# Selective area epitaxy of PbTe-Pb hybrid nanowires on a lattice-matched substrate

Yuying Jiang[1]*, Shuai Yang[2]*, Lin Li[2]*, Wenyu Song[1]*, Wentao Miao[1]*, Bingbing Tong[2], Zuhan Geng[1], Yichun Gao[1], Ruidong Li[1], Qinghua Zhang[3], Fanqi Meng[3], Lin Gu[3], Kejing Zhu[2], Yunyi Zang[2], Runan Shang[2], Xiao Feng[1,2,4], Qi-Kun Xue[1,2,4,5], Dong E. Liu[1,2,4], Hao Zhang[1,2,4]†, and Ke He[1,2,4]‡

[1]*State Key Laboratory of Low Dimensional Quantum Physics, Department of Physics, Tsinghua University, Beijing 100084, China*

[2]*Beijing Academy of Quantum Information Sciences, Beijing 100193, China*

[3]*Institute of Physics, Chinese Academy of Sciences, Beijing 100190, China*

[4]*Frontier Science Center for Quantum Information, Beijing 100084, China*

[5]*Southern University of Science and Technology, Shenzhen 518055, China*

*\* These authors contributed equally to this work.*

† *hzquantum@mail.tsinghua.edu.cn*

‡ *kehe@tsinghua.edu.cn*

**Topological quantum computing is based on braiding of Majorana zero modes encoding topological qubits. A promising candidate platform for Majorana zero modes is semiconductor-superconductor hybrid nanowires. The realization of topological qubits and braiding operations requires scalable and disorder-free nanowire networks. Selective area growth of in-plane InAs and InSb nanowires, together with shadow-wall growth of superconductor structures, have demonstrated this scalability by achieving various network structures. However, the noticeable lattice mismatch at the nanowire-substrate interface, acting as a disorder source, imposes a serious obstacle along with this roadmap. Here, combining selective area and shadow-wall growth, we demonstrate the fabrication of PbTe-Pb hybrid nanowires—another potentially promising Majorana system—on a nearly perfectly lattice-matched substrate CdTe, all done in one molecular beam epitaxy chamber. Transmission electron microscopy shows the single-crystal nature of the PbTe nanowire and its atomically sharp and clean interfaces to the CdTe substrate and the Pb overlayer, without noticeable inter-diffusion or strain. The nearly ideal interface condition, together with the strong screening of charge impurities due to the large dielectric constant of PbTe, hold promise towards a clean nanowire system to study Majorana zero modes and topological quantum computing.**

Majorana zero modes (MZMs)[1,2] have attracted enormous interests in the past decade for their non-Abelian statistics and potential applications in topological quantum computing (TQC)[3,4]. As a promising platform to realize MZMs and TQC[5,6], semiconductor-superconductor hybrid nanowires have been extensively studied since 2010 with sustained efforts from theory prediction, to material growth, device fabrication and transport measurement, mostly in InAs- and InSb-based systems[7-9]. Guided by theory[10,11] and enabled by material growth[12-14], electron transport experiments on hybrid nanowires have been significantly advanced[15-22] where to date, however, an unambiguous demonstration of MZMs is still lacking. Recent theory developments point out that disorder, even after years of optimization for ballistic transport[23-25], still plays a major role in currently the best InAs/InSb hybrid nanowire devices[26-28]. One solution to the problem is to switch to superconductors with larger superconducting gaps, e.g. Pb[29] and Sn[30], which presumably have higher tolerance of disorder than the commonly used Al. Another solution, seemingly more difficult, is to replace InAs/InSb with another semiconductor less influenced by disorder[31-33].

In this paper, we demonstrate the molecular beam epitaxy (MBE) growth of PbTe-Pb hybrid nanowires, a promising material candidate for Majorana search with improvement in both the superconductor and semiconductor respects. PbTe is a IV-VI semiconductor with strong spin-orbit coupling and a bandgap and Lander g factor similar to InAs/InSb. The ultra-large dielectric constant of PbTe, two orders of magnitude higher than InAs and InSb, can significantly screen charge disorder[34]. Early experiments on PbTe-based two dimensional electron gas have shown ballistic transport[35-38], quantum Hall effect[39-42] and mobility larger than $2\times10^5$ cm$^2$/Vs[42]. Lead can induce a proximity superconducting gap in PbTe-family materials much larger than Al in InAs/InSb[43,44]. Isolated MZMs are predicted to appear in PbTe-Pb nanowires with the valley degeneracy of PbTe lifted by certain crystalline orientations and device geometries[34].

Moreover, one can find an ideal substrate—CdTe—for selective area growth (SAG) of in-plane epitaxial PbTe nanowires and networks. Previously reported SAG InAs[45-50] and InSb[51-53] nanowires, though demonstrating the scalability needed for future braiding circuits, suffer from the lack of lattice-matched substrates. The lattice mismatch, a serious disorder source, significantly reduces the quality of SAG nanowires and is considered as the biggest barrier for the SAG approach to scalable topological quantum computing. CdTe is a commonly used substrate for growing II-VI and IV-VI semiconductors with a bandgap of 1.5 eV and type I band alignment with PbTe. PbTe and CdTe have nearly identical lattice constants but distinct crystalline structures (rocksalt for PbTe and zincblende for CdTe), which minimizes both the strain and intermixing in their heterostructures[54,55]. Therefore, the PbTe nanowires epitaxied on CdTe substrates are expected to show an ideal interface.

In this study, we not only grow epitaxial PbTe nanowires on CdTe substrates with SAG, but also *in situ* prepare Pb nano-structures on the PbTe nanowires using shadow-wall growth in the same MBE chamber [14,25,56-59], avoiding the uncontrollable Pb etching process that may deteriorate the device quality. Our results provide a complete toolbox of fabricating scalable topological quantum computation devices with a new (potentially better) hybrid nanowire system for Majorana detection.

Figure 1 shows the fab/growth sequence schematic of the PbTe-Pb hybrid nanowires (see Fig. 1a the crystal structure). We start with a commercial CdTe(001) substrate (green in Fig. 1b) on which silicon oxide (SiO$_x$) shadow walls (dark grey) are fabricated by electron beam lithography (EBL), see SFig. 1 for details. The shadow walls, 500 – 700 nm high, are made from inorganic negative resist Hydrogen SilsesQuioxane (HSQ) which is converted into SiO$_x$ after electron beam exposure. A 20 – 40 nm thick Al$_2$O$_3$ mask layer (light grey in Fig. 1c) is then uniformly deposited by magnetron sputtering, followed by a wet etching, using Transene Etchant Type D, to open nanowire-shape trenches (white line in Fig. 1d) for SAG. After cleaning by oxygen plasma, the pre-patterned substrates are loaded into the MBE chamber (Omicron Lab-10), in which they are further treated by Ar$^+$ sputtering and annealing at 280°C until clean CdTe surface at the mask openings is obtained.

PbTe nanowires are grown with a PbTe compound source (99.999%) evaporated by a standard Knudson cell with the beam flux ~1.24 ML/min. Selective growth of PbTe on the exposed CdTe surface at the mask openings is achieved (Fig. 1e) when the substrate temperature ($T_{\text{subs}}$) is between ~310 °C and ~340 °C. To grow Pb nanostructures on the SAG PbTe nanowires, the sample is *in situ* transferred to another sample stage in the same MBE chamber. Here, the sample is cooled to ~ 100 K with liquid nitrogen, as required by 2D growth of Pb films, and tilted such that the angle between the incident Pb flux and sample surface normal direction is 54° – 61°. Such an evaporation geometry and the 500-700 nm high shadow walls result in a Pb film with 1 – 1.3 μm long shadows grown on the substrate (Fig. 1f). A CdTe capping layer (Fig. 1g) is grown to cover the nanowire sample before it is taken out for further fabrications and measurements.

Figure 2a shows the scanning electron microscope (SEM) image of a PbTe nanowire (along the [010] direction) grown on a CdTe substrate without shadow walls. The nanowire is uniform with a flat surface and nearly straight edges with the length exceeding 7 μm and the width below 200 nm. The atomic force microscope (AFM) image on the surface of such a nanowire (SFig. 2) shows atomically flat terraces spaced by ~3.23 Å high steps, corresponding to the interlayer spacing of PbTe (001). The step-terrace structure is typical of an epitaxial film on a single crystal substrate. More complex nanowire-based structures are obtained in mask openings of different shapes, such as the loop shown in Fig. 2b and the double crosses shown in Fig. 2c, both composed of nanowires of <100> directions. The two kinds of structures can be used in future Aharonov-Bohm (AB) effect and Hall effect experiments, respectively. These structures share the same uniformity as the nanowires without obvious faceting observed at crossings and corners, since {001} family surfaces of rocksalt crystals have the lowest surface energy. PbTe nanowires along <110> directions with similar uniformity can also be obtained, though requiring more careful tuning of the growth parameters. Faceted edges appear in the nanowires along other directions (SFig. 3).

Figures 2d-f display the SEM images of three different PbTe-Pb in-plane junctions prepared by shadow wall growth of Pb on SAG PbTe nanowires (along the [110] direction). The three structures, made by different positions of shadow walls (bright line shapes) relative to PbTe nanowires, correspond to N-S, S-S and N-S(island)-N devices, respectively (N for normal metal and S for superconductor). N-S devices can be used for tunnelling spectroscopy to study the induced superconducting gap and Majorana or Andreev sub-gap states. S-S devices are for Josephson junctions while N-S(island)-N devices for studies of 'Majorana island'. The Pb overlayers grown on PbTe nanowire are atomically flat as shown in the AFM image (SFig. 2). On the $Al_2O_3$ mask, the Pb layer is much rougher, as a result of merging of 3D Pb islands with different crystalline orientations. The edges of the Pb overlayers on the PbTe surface are very abrupt, making nearly ideal termination of the superconducting region hosting MZMs.

High-resolution scanning transmission electron microscopy (STEM) was used to characterize the cross-sectional structure of a PbTe-Pb hybrid nanowire along the [110] direction (56 nm PbTe nanowire with a 16-18 nm thick Pb film and 10 nm CdTe capping layer). Figure 3a shows the high-angle annular dark field (HAADF) image, from which the CdTe substrate, the PbTe wire, the Pb overlayer, and the CdTe capping layer are clearly distinguished. The PbTe wire is partly grown in the trench of CdTe substrate etched by $Ar^+$ sputtering and partly overgrown out of the $Al_2O_3$ mask. The overgrown can be controlled by adjusting the growth time. The Pb overlayer becomes blurred at the edges, presumably broken up by the focused ion beam (FIB) process during the TEM sample preparation (note that the top CdTe capping layer is kept intact, suggesting that the breakup of the Pb layer occurred after the growth was completed). Despite these imperfections, the interfaces between neighbouring layers are rather sharp, suggesting little inter-layer diffusion. It is further confirmed by the energy-dispersive x-ray spectroscopy (EDX) maps of Pb, Cd and Te in Figs. 3b, 3c and 3d, respectively. The characteristic signals of each element change abruptly across the interfaces to neighbouring layers (SFig. 4). The atomically resolved STEM images around the CdTe-PbTe interface is shown in Fig. 3e and SFig. 5. The PbTe nanowire is nearly perfectly epitaxied on the CdTe substrate with an atomically abrupt interface. From the inverse fast Fourier transformation (IFFT) of the TEM image in Fig. 3f, we observe no misfit dislocation which widely exists in InSb-InP and InAs-InP interfaces in early works. The PbTe-Pb interface is also atomically sharp as shown in Fig. 3g. From the measured interlayer spacing of ~2.85 Å, we conclude that the Pb overlayer is (111)-oriented. The formation of the (111) oriented Pb layer on PbTe(001) surface is due to the low surface energy of Pb(111) surface and the weak bonding between PbTe and Pb. The different surface symmetries rule out any epitaxial relationship between the PbTe(001) nanowire and the Pb(111) overlayer. Nevertheless, no grain boundaries are resolved in the Pb(111) overlayer, suggesting that the size of Pb grains is at least larger than the width of the nanowire. The CdTe capping layer grown on the Pb layer is multi-crystalline (Fig. 3h), but it also forms an atomically abrupt interface with the Pb layer.

In Fig. 4, we display the preliminary electron transport characterization of a SAG PbTe device. Figure 4a shows the device SEM image: a PbTe nanowire is contacted by two normal metal electrodes (Ti/Au) after a short *in situ* Ar plasma etching for Ohmic contacts. The channel length (contact spacing) of the device is ~2.8 μm, long enough for diffusive electron transport. After the measurement, the device was further cut for TEM and EDX analysis (SFig. 6) with its cross-section shown in Fig. 4b. Figure 4c shows the device conductance tuned by voltage applied to a top gate ($V_{TG}$). The channel can be fully pinched off and opened as a field-effect transistor. We notice a sizable hysteresis of the pinch off curves for different $V_{TG}$ sweeping directions. The hysteresis is reduced by an order of magnitude in PbTe-Pb hybrid nanowires based on our preliminary characterization, therefore does not impose a serious issue for future superconductivity related studies. For the upward sweeping, we can fit the curve to extract the device mobility based on the formula[23]: $G(V_{TG}) = \left(R_c + \frac{L^2}{\mu C(V_{TG}-V_{th})}\right)^{-1}$, with $L$ representing channel length, $\mu$ carrier mobility, $V_{th}$ threshold voltage, $R_C$ the series resistance, and $C$ gate-nanowire capacitance. A key parameter of the fitting is the capacitance $C$, which we estimate to be 0.709 fF based on a Laplace Solver using finite element method by taking the device geometry (Fig. 4b) as an input[23]. Quantum confinement effect is also considered which further reduces the capacitance by ~ 20%[60]. The fitting curve (red line) shows reasonable agreement with the data (black) with an extracted field effect mobility ~ $1.5 \times 10^4$ cm$^2$/Vs, a value comparable (same order of magnitude) to those of InSb nanowires[23]. During the growth of this device, the PbTe nanowire was capped by CdTe (SFig. 6) to protect the device from possible surface disorder. Without this capping layer (see SFig. 7), the device mobility is significantly reduced (roughly an order of magnitude), suggesting the necessity of lattice-matched capping for high quality devices. For the downward sweeping curve, we could not find a reasonable fit using this mobility model. Figure 4d shows the magneto-conductance of the device where a conductance peak near zero magnetic field indicates the effect of weak anti-localization due to the strong spin-orbit coupling of the PbTe nanowire. Systematic transport studies on mobility optimization, WAL (to extract spin-orbit interaction), AB interference (to extract phase coherence length) and superconductivity related physics (in PbTe-Pb hybrids) are beyond the scope of this (growth) paper and will be addressed in our future works.

From the characterization results shown above, we can see that although the ultra-high-vacuum (UHV) in-situ fabricated PbTe-Pb devices are still far from perfect, they do solve several key problems of 'Majorana nanowires': scalability of nanowire networks, lattice match with the substrate, and atomically abrupt and clean interfaces. The large dielectric constant of PbTe is expected to significantly reduce the potential variation induced by charge impurities. The insensitivity to disorder probably explains the mobility of ~$10^4$ cm$^2$/Vs achieved in the preliminary devices presented in this work (In total, we have successfully fabricated and measured three mobility devices shown in Fig. 4 and SFig. 7 for the first round).

With future optimization of growth and device fabrication (e.g. mask uniformity, substrate processing, growth condition, single-nucleation, etc), we expect higher mobility to be achieved. Actually, the mobility over $10^6$ cm$^2$/Vs has already been reported in PbTe films due to the charge screening effect of the large dielectric constant[61], despite far fewer studies devoted to the material than to III-V semiconductors. Therefore, PbTe nanowires have a big advantage in overcoming the disorder problem, the most formidable one now in nanowire-based approach to topological quantum computing.

In summary, we have successfully prepared various PbTe-Pb hybrid nanowires on lattice-matched CdTe substrates by combining SAG and shadow wall growth in the same MBE chamber. The PbTe nanowires have carrier mobility up to ~$10^4$ cm$^2$/Vs and form atomically sharp and clean interfaces with the CdTe substrate and Pb overlayer. With the disorder induced by strain and interfaces significantly suppressed and the impurity charges screened by the large dielectric constant, PbTe-Pb hybrid nanowires provide a clean platform to study MZMs and related physics. The UHV device fabrication techniques of the hybrid PbTe-Pb nanowire system developed in this work pave the road towards scalable and disorder-free Majorana networks and circuits to realize topological qubits and braiding.

**Acknowledgement** We thank Leo Kouwenhoven and Erik Bakkers for valuable discussions. We also thank Han Ying (School of Integrated Circuits, Tsinghua University) for technical assistance. During the preparation of this manuscript, we became aware of a similar work from Bakkers group.

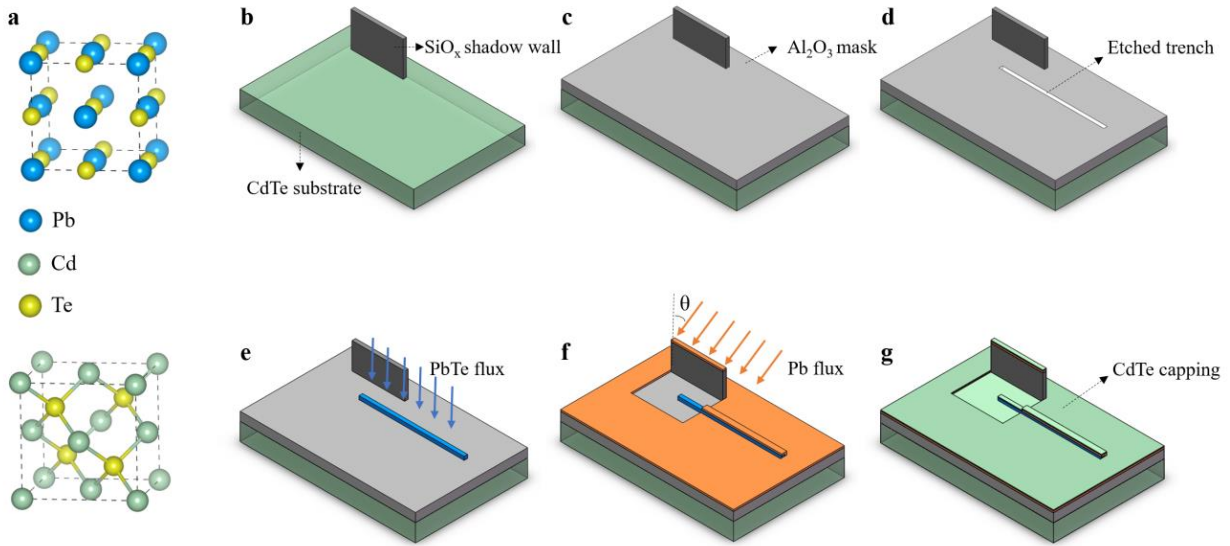

**Figure 1 | Schematic fabrication procedure of pre-patterned CdTe substrates and PbTe-Pb hybrid nanowires. a** Crystal structure of PbTe and CdTe. **b** Preparation of SiO$_x$ shadow walls (dark grey) on a CdTe substrate (green). **c** Al$_2$O$_3$ mask (light grey) deposition. **d** Wet etching of Al$_2$O$_3$ with openings on a CdTe substrate (white). **e** Selective area growth of PbTe nanowires (blue). The arrows represent the PbTe flux direction. **f** Shadow wall growth of Pb overlayer (orange). The arrows represent Pb flux direction. **g** Capping the device with CdTe.

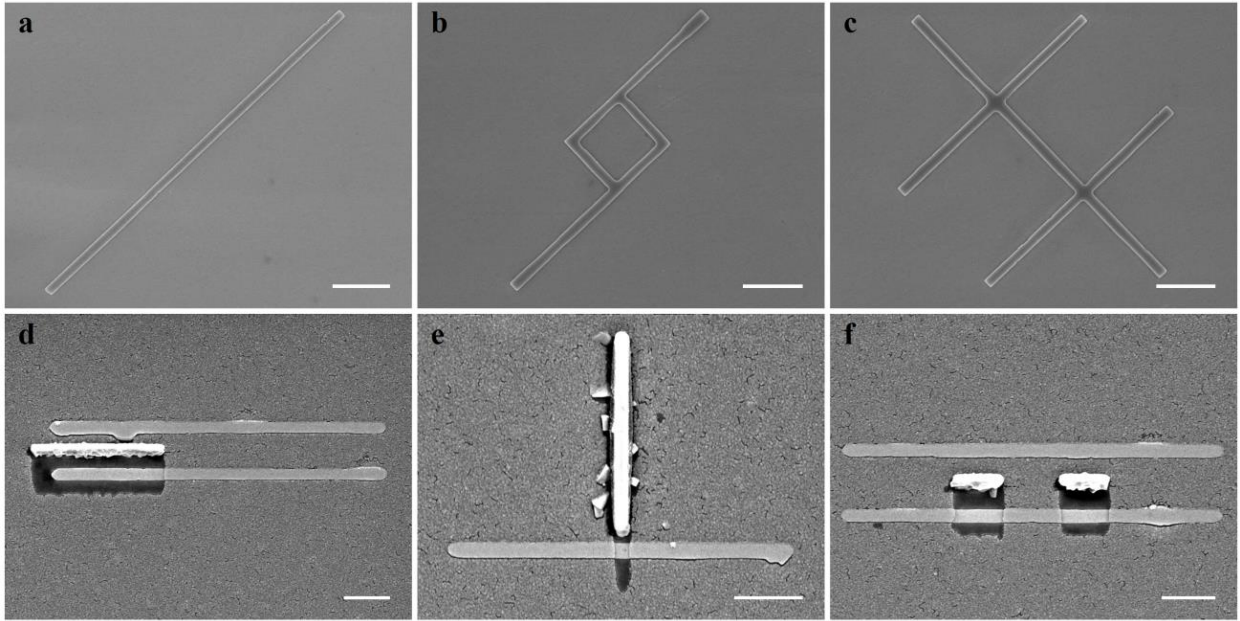

**Figure 2 | Scanning electron microscope (SEM) images of PbTe and PbTe-Pb nanowires and nanostructures grown on CdTe(001). a-c** SEM images of a PbTe nanowire (a), an Aharonov-Bohm loop (b) and a double cross (Hall bar) composed of PbTe nanowire (c), all selectively grown on CdTe (001). **d-f** SEM images of PbTe-Pb hybrid nanowire structures for NS tunnelling spectroscopy (d), SS Josephson junction (e) and superconducting island devices (f), respectively. All the scales bars are 1 micron.

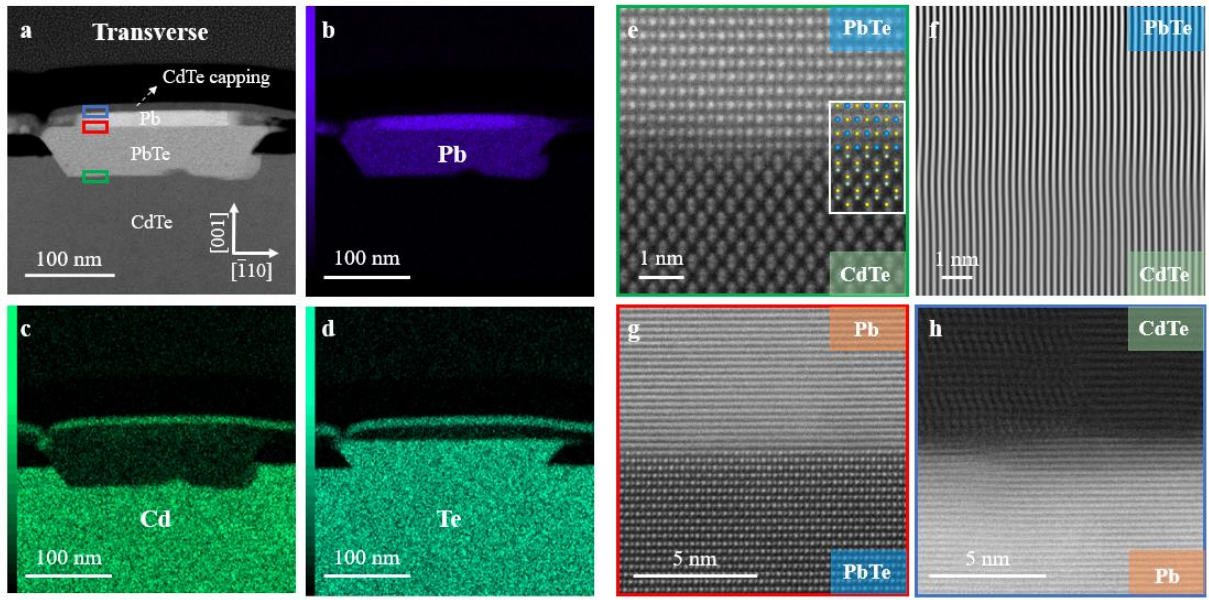

**Figure 3 | Scanning transmission electron microscopy characterizations of the cross-section of a PbTe-Pb hybrid nanowire grown on CdTe(001) substrate.** **a** High-angle annular dark field (HAADF) image. **b-d** Energy-dispersive x-ray spectroscopy (EDX) maps of Pb (b), Cd (c) and Te (d), respectively. **e** Atomically resolved image near the CdTe-PbTe interface. **f** Inverse fast Fourier transformation (IFFT) of e, resolving the columns of atoms as vertical lines. **g, h** Atomically resolved images near the PbTe-Pb interface (g) and CdTe capping layer-Pb interface (h).

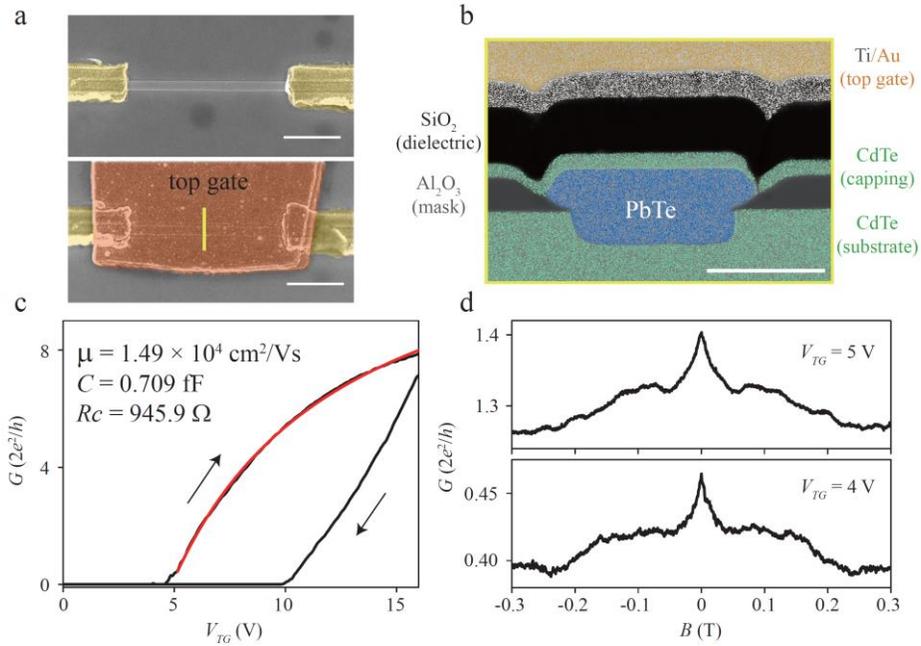

**Figure 4 | Basic transport characterization of PbTe nanowires. a**, False-colour SEM of a field effect mobility device (Device A). For clarity, the upper (lower) panel is the device without (with) dielectric and top gate (orange). Scale bar is 1 micron. **b**, Cross-section of the device (after measurement) with each layer labelled. Scale bar is 100 nm. **c**, Conductance of the device as a function of top gate voltage ($V_{TG}$). The arrows indicate $V_{TG}$ sweep directions. The red line is the mobility fitting curve. **d**, Magneto-conductance of the device at two $V_{TG}$ settings, resolving a weak anti-localization peak near $B = 0$ T. The magnetic field ($B$) direction is perpendicular to the substrate plane. All measurements are formed in a dilution fridge with a base temperature ~ 20 mK.

# Supplement Information: Selective area epitaxy of PbTe-Pb hybrid nanowires on a lattice-matched substrate

Yuying Jiang[1]*, Shuai Yang[2]*, Lin Li[2]*, Wenyu Song[1]*, Wentao Miao[1]*, Bingbing Tong[2], Zuhan Geng[1], Yichun Gao[1], Ruidong Li[1], Qinghua Zhang[3], Fanqi Meng[3], Lin Gu[3], Kejing Zhu[2], Yunyi Zang[2], Runan Shang[2], Xiao Feng[1,2,4], Qi-Kun Xue[1,2,4,5], Dong E. Liu[1,2,4], Hao Zhang[1,2,4†], and Ke He[1,2,4‡]

[1]*State Key Laboratory of Low Dimensional Quantum Physics, Department of Physics, Tsinghua University, Beijing 100084, China*

[2]*Beijing Academy of Quantum Information Sciences, Beijing 100193, China*

[3]*Institute of Physics, Chinese Academy of Sciences, Beijing 100190, China*

[4]*Frontier Science Center for Quantum Information, Beijing 100084, China*

[5]*Southern University of Science and Technology, Shenzhen 518055, China*

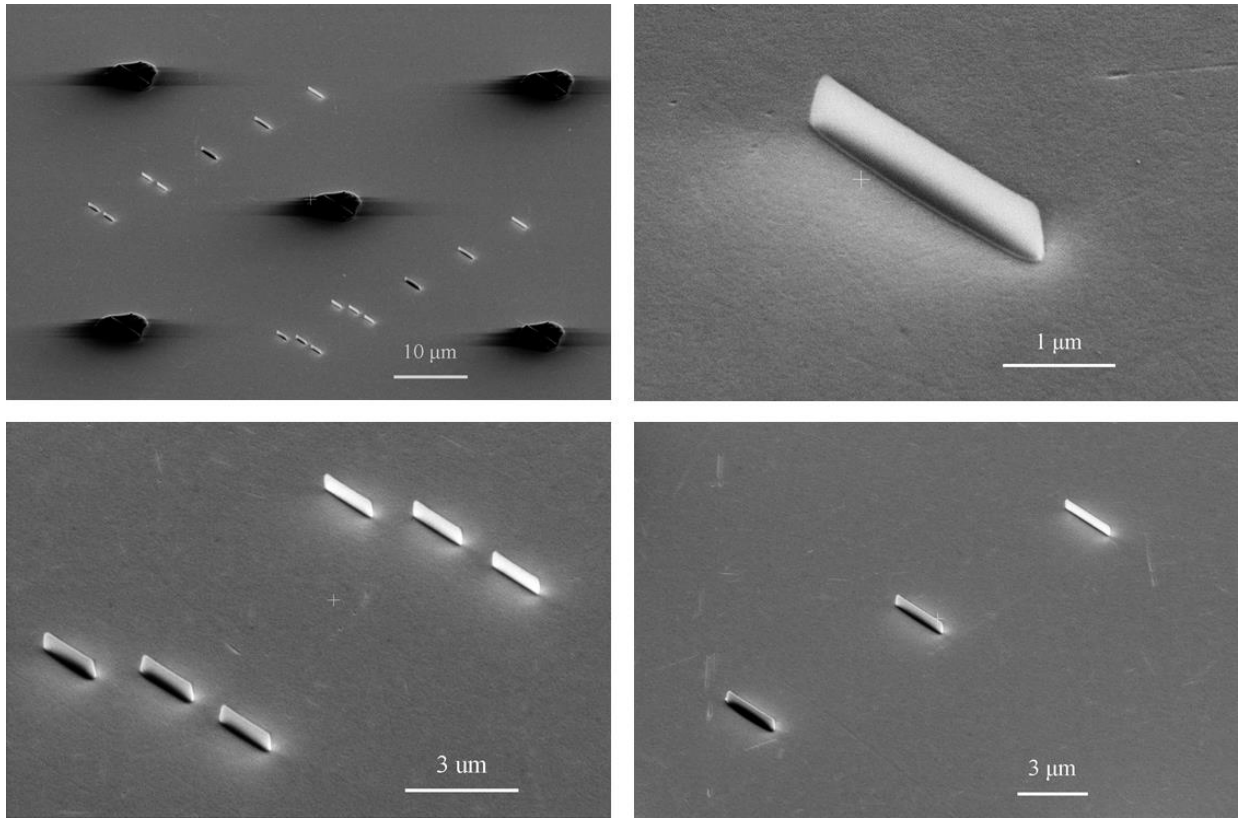

**Supplement Figure 1 | Tilted SEM images of the SiOx shadow walls (bright).** The five big 'black dots' in the upper left panel are HSQ markers.

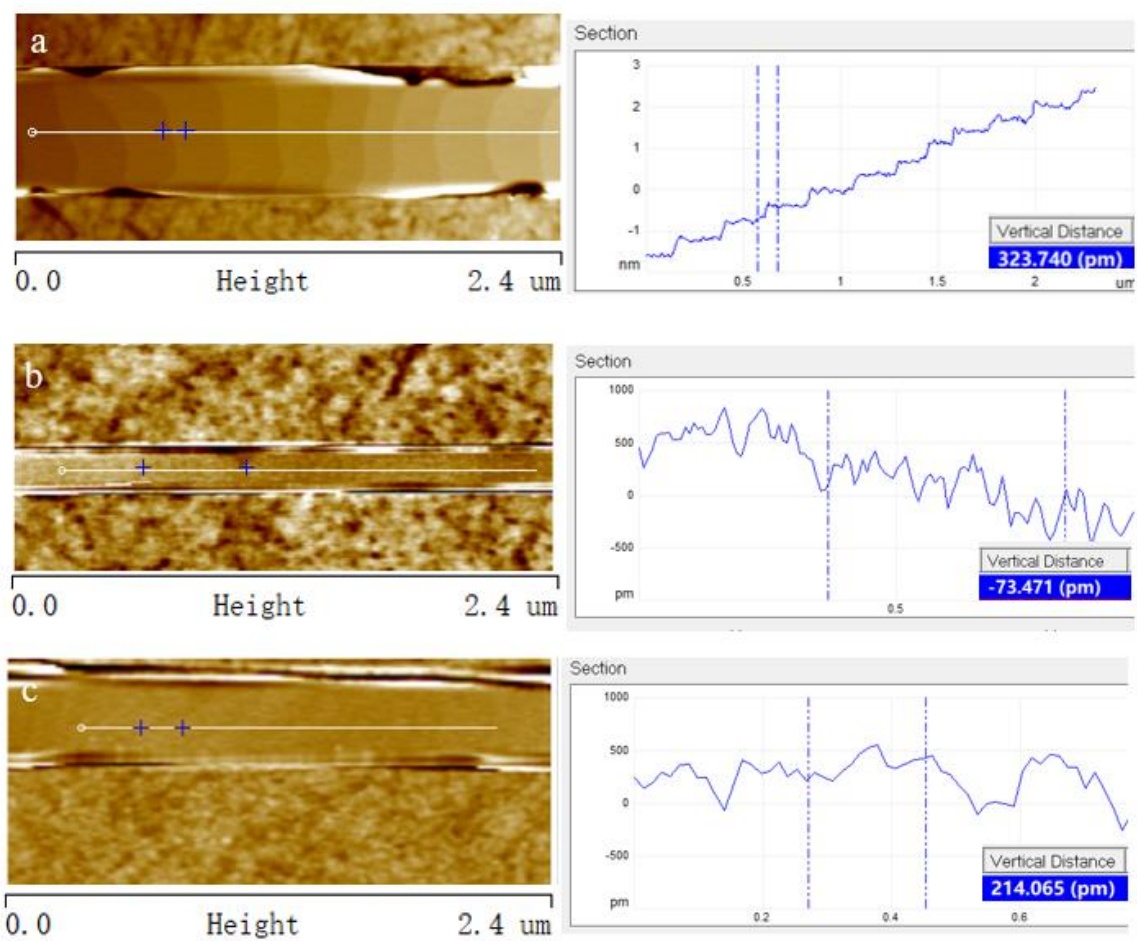

**Supplement Figure 2 | AFM images showing morphologies of the PbTe nanowire, CdTe/PbTe and Pb/PbTe hybrids. a** Atomically flat morphology of a PbTe nanowire along [010] direction. The height of adjacent steps is about 3.23Å, corresponding to the interlayer spacing of PbTe (001). **b** The morphology of a PbTe nanowire capped with CdTe at liquid nitrogen temperature. **c** The morphology of Pb on a PbTe-Pb hybrid nanowire. The overall fluctuation is within 0.5 nm.

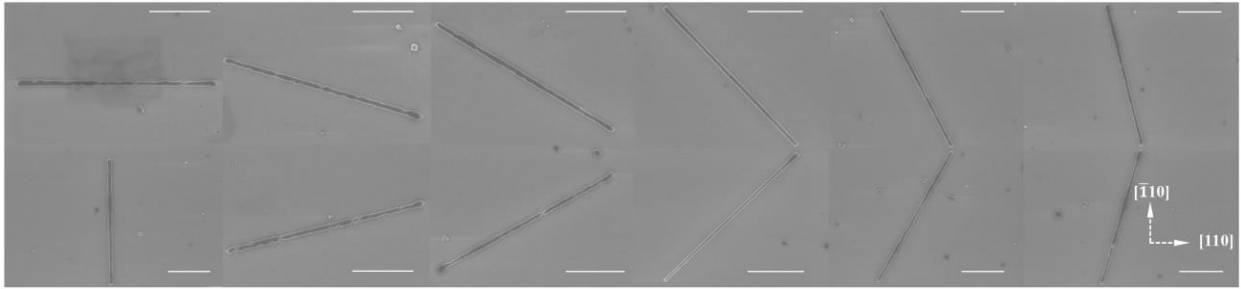

**Supplement Figure 3 | SEM images of PbTe nanowires grown along different crystal directions.** All the nanowires are ~ 7 μm in length and roughly 150 nm in width. All the scale bars are 2 microns. The rectangular CdTe substrate edges are along the family of <110> directions. We design a set of mask openings rotated every 15 degrees from the substrate edge. At the given growth temperature (320℃) and growth time (3.5 hours at the rate of 1.2 ML/min), the best uniformity is achieved on PbTe nanowires along <100> directions. This is expected since the {100} family of surfaces of rocksalt crystals have the lowest surface energy. By increasing the growth time to 4.5 hours, better uniformity can be achieved on PbTe nanowires along <110> directions, see Fig 2 (d-f).

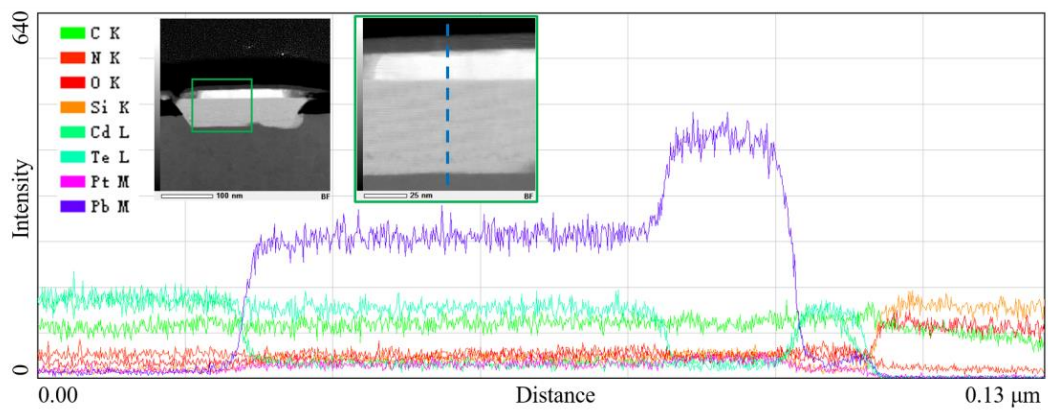

**Supplement Figure 4 | EDX line analysis showing characteristic signals of each element across interlayers.** The inset figures are cross-section images of the lamella cut from a hybrid PbTe-Pb nanowire, which is the same sample as shown in Fig 3. EDX line scan is performed along the blue dashed line from bottom to top. Characteristic signals of each element change abruptly across the interfaces to neighbouring layers.

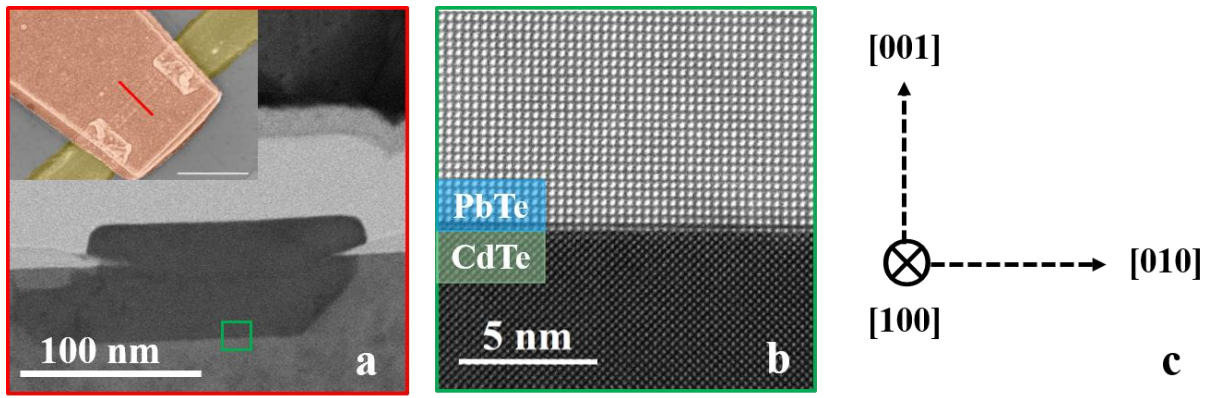

**Supplement Figure 5 | STEM characterizations of the cross-section of a PbTe nanowire grown along [100] direction, cut from device C. a** HAADF image of the cross-section. The inset figure shows an SEM image of the device, where the cross-section is indicated by the red line. **b** Atomically resolved image near the CdTe-PbTe interface, looking along the [100] direction. The interface and PbTe crystal lattice have the same high quality as the nanowire along [110] directions, see Fig 3. **c** Notification of the crystal directions in b.

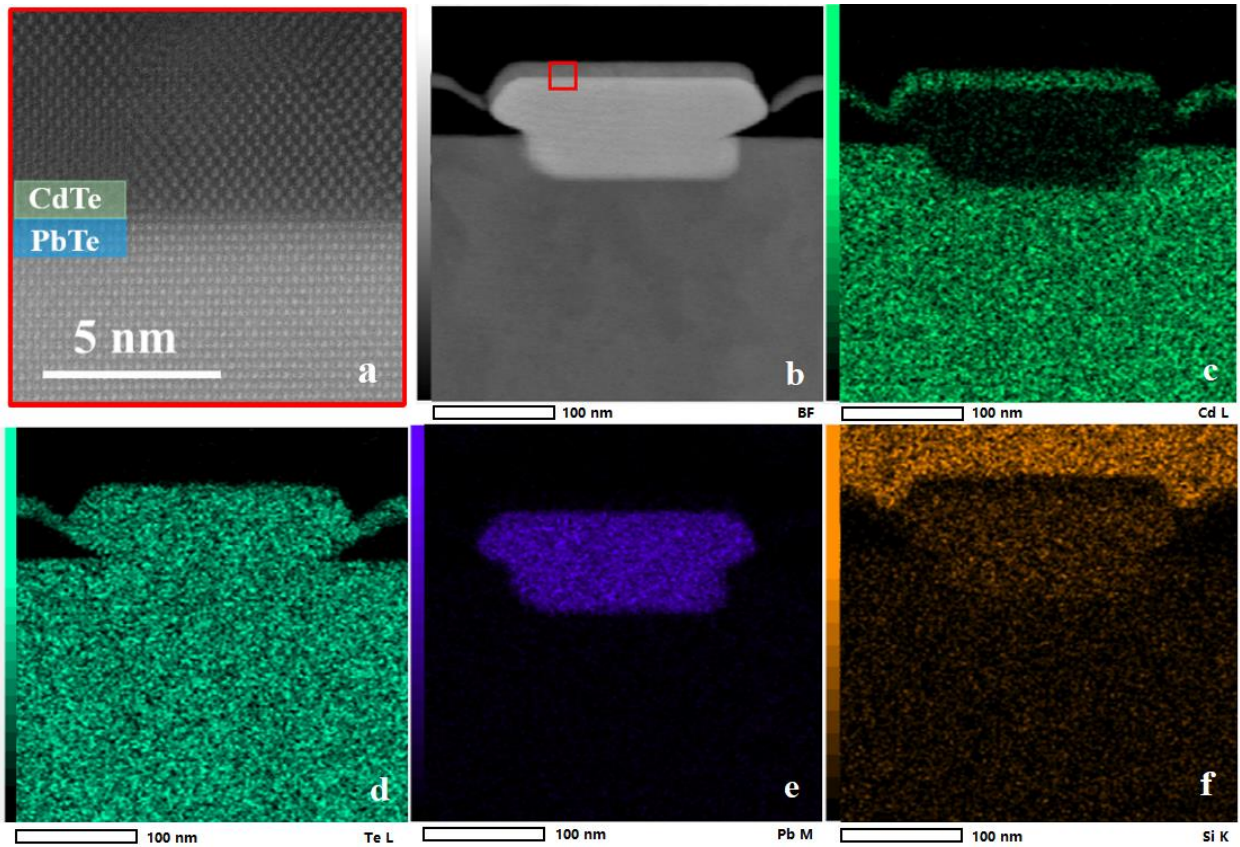

**Supplement Figure 6 | Scanning transmission electron microscopy (STEM) characterizations of the PbTe nanowire cross-section, cut from device A. a** Atomically resolved image near the PbTe-CdTe capping layer interface. The interface is atomically flat without any signs of interdiffusion. CdTe capping layer is poly-crystalline probably because it was grown at low temperature ~ 100 K. **b** High-angle annular dark field (HAADF) image of the cross-section. **c-f** Energy-dispersive x-ray spectroscopy (EDX) maps of Cd (c), Te (d), Pb (e) and Si (f) elements respectively.

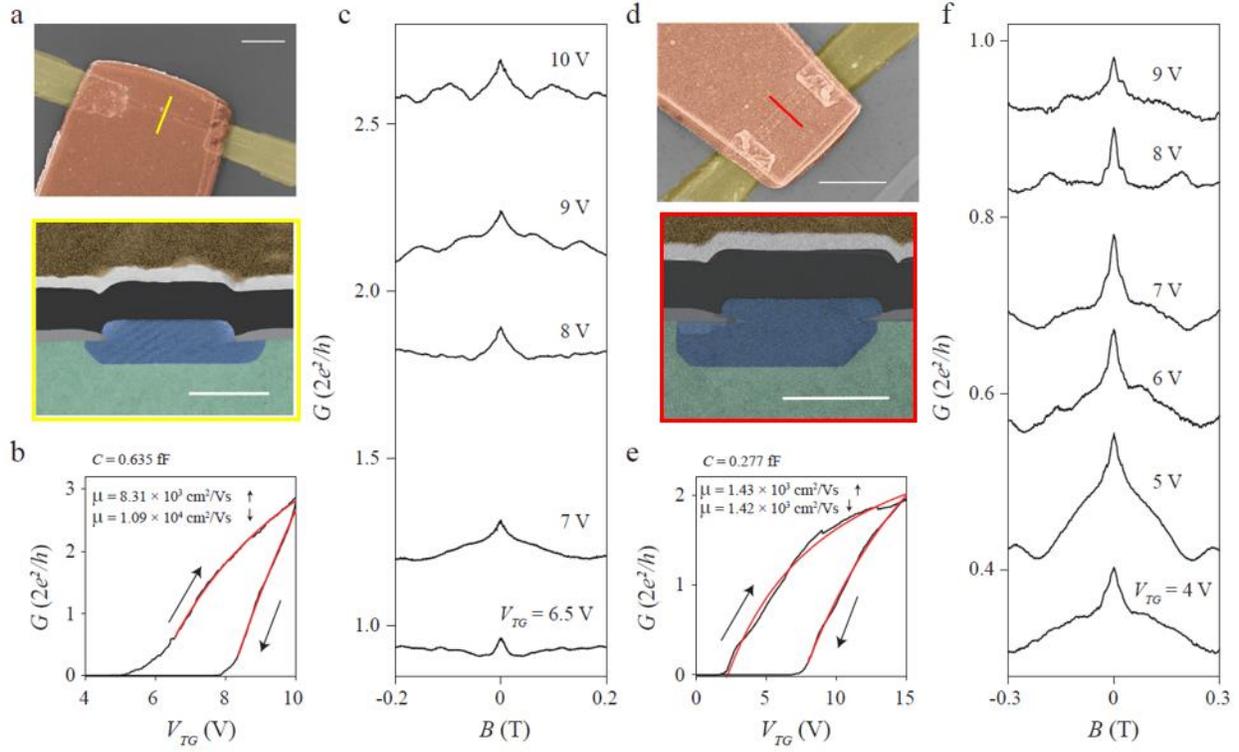

**Supplement Figure 7 | Electron transport characterization of two more SAG PbTe nanowire devices.** **a-c** for Device B (PbTe capped with ~ 6 nm thick CdTe at 270 °C) and **d-f** for Device C (no CdTe capping). **a, d** show the device SEM and cross-section geometry. **b, e** show the pinch off curves and mobility fit (red). The extracted mobility is labelled for the upward and downward sweeping (arrows). Based on TEM, the CdTe capping is probably not continuous (fully capping) on the PbTe surface. **c, f** show the magneto-conductance scan at various difference gate voltages, resolving the weak anti-localization peak.